\magnification1200


\def\title{
\centerline{Monopole supersymmetries }
\smallskip\centerline{and }\smallskip
\centerline{the Biedenharn operator}
}

\def\runningtitle{Monopolesusy}

\def\author{\vfill
\centerline{P. A. HORV\'ATHY\foot{
Permanent address: Laboratoire de Math\'ematiques et
de Physique Th\'eorique, Universit\'e de TOURS.
F-37 200 TOURS (France).
 e-mail: horvathy@univ-tours.fr},
A. J. MACFARLANE\foot{Permanent address: Centre for Mathematical Sciences,
DAMTP, University of Cambridge,
Wilberforce Road, Cambridge CB3 0WA, UK. \hfill\break
e-mail: a.j.macfarlane@damtp.cam.ac.uk},
J.-W. VAN HOLTEN\foot{e-mail : t32@nikhef.nl}
}
\vskip 3mm
\centerline{NIKHEF, Postbus 41882}\smallskip
\centerline{1009 DB Amsterdam, The Netherlands}
\vfill
}

\def\runningauthor{
Horv\'athy, Macfarlane, van Holten
}


\vsize = 9.2truein
\hsize = 6.4truein 
\baselineskip = 12 pt 

\headline ={
\ifnum\pageno=1\hfill
\else\ifodd\pageno\hfil\tenit\runningtitle\hfil\tenrm\folio
\else\tenrm\folio\hfil\tenit\runningauthor\hfil
\fi\fi
} 

\nopagenumbers
\footline = {\hfil} 


\def\parag{\hfil\break} 
\def\IR{{\bf R}} 
\def\ccr{\cr\noalign{\medskip}}
\def\smallcirc{{\raise 0.5pt\hbox{$\scriptstyle\circ$}}}
\def\and{\qquad \hbox{and}\qquad}

\def\2{{1\over 2}}

\def\smallover#1/#2{\hbox{$\textstyle{#1\over#2}$}} %
\def\2{{\smallover 1/2}}
\def\parag{\hfil\break} 
\def\={\!=\!}
\def\D{{D\mkern-2mu\llap{{\raise+0.5pt\hbox{\big/}}}\mkern+2mu}\ }
\def\osp{{\rm osp}}
\def\o{{\rm o}}
\def\kikezd{\parag\underbar}


\newcount\ch 
\newcount\eq 
\newcount\foo 
\newcount\ref 

\def\chapter#1{
\parag\eq = 1\advance\ch by 1{\bf\the\ch.\enskip#1}
}

\def\equation{
\leqno(\the\ch.\the\eq)\global\advance\eq by 1
}

\def\foot#1{
\footnote{($^{\the\foo}$)}{#1}\advance\foo by 1
} 

\def\reference{
\parag  [\number\ref]\ \advance\ref by 1
}

\ch = 0 
\foo = 1 
\ref = 1 

\title
\vskip 5mm
\author
\vskip 2mm
\parag{\bf Abstract.}
{\it The hidden supersymmetry of the monopole found
by De Jonghe et al.  is generalized to
 a spin $\2$ particle in the
combined field of a Dirac monopole plus a $\lambda^2/r^2$
potential [considered before by D'Hoker and Vinet],
and related to the operator introduced by Biedenharn a long time ago
in solving the Dirac-Coulomb problem.
Explicit solutions are obtained by diagonalizing the
Biedenharn operator}.
\vskip4mm

\chapter{Introduction}

In  [1], De Jonghe et al. found that a spin-$\2$ particle
of mass, here chosen $m=\2$,
and charge $q$ in the field of a Dirac monopole of unit strength,
described by the Pauli Hamiltonian
$$
H_D=\Big(\vec{D}^2-q{\vec\sigma\cdot\vec{r}\over r^3}\Big),
\equation
$$
admitted a `hidden' supercharge $\widetilde{Q}_D$ given by
$$
\widetilde{Q}_D=\big(\vec\sigma\cdot\vec\ell+1\big),
\equation
$$
where
$\vec\ell$ is the [non-conserved] orbital part of the angular momentum,
$\vec\ell=\vec{r}\times\vec\pi$, $\vec\pi=\vec{p}-q\vec{A}$.
Their discussion is based on the study of Killing-Yano tensors.
The supercharge
 $\widetilde{Q}_D$ anticommutes with the supercharge $Q_D$ found by
D'Hoker and Vinet [2], where
$$
Q_D=\vec\sigma\cdot\vec\pi.
\equation
$$
These supercharges form, together with the [conserved]
total angular momentum \hfill\break
$
\vec{J}\=\vec{ L}+{\vec\sigma\over2},
\;
\vec{ L}\=\vec{ \ell}- q{\vec{ r}/r},
$
a closed, non-linear algebra  [1].

In this Letter, we
(i) generalize the result of De Jonghe et al.;
(ii)  relate it to earlier work of
Dirac [3], Biedenharn [4], and Berrondo and McIntosh [5];
(iii) use it to solve the system, and
(iv) discuss a full Minkowski space generalisation.

Very recently  [6], Plyushchay
discussed related  problems, but
from a rather different veiwpoint~:
while our results here are derived from
supersymmetric quantum  mechanics,
he uses pseudoclassical mechanics
with anticommuting (Grassmann) variables.
See also the comments on [1] in [7].
\goodbreak

\chapter{The generalized monopole system}

Let $\vec{ A}_D$ denote the vector potential of the Dirac monopole
of unit strength.
Let $\lambda$ be a positive constant, $q>0$ a  half-integer,
and consider, setting
$A_4\equiv\lambda/r$ and
$\vec{ A}\equiv q\vec{ A}_D$,
the gauge field $A_\alpha$, ($\alpha=1,\dots,4$) on $\IR^4$.
We use the Dirac matrices
$$
\vec{\gamma}=
\pmatrix{&-i\vec{\sigma}\cr-i\vec{\sigma}&\cr},
\quad
\gamma^4=
\pmatrix{&{\bf1}_2\cr-{\bf1}_2&\cr},
\gamma^5=
\pmatrix{{\bf1}_2&\cr&-{\bf1}_2\cr}.
\equation
$$
When restricted to fields which do not depend on $x^4$,
the associated Dirac operator $\D\equiv\gamma^\mu D_\mu$,
$D_\mu\equiv\partial_\mu-iA_\mu$,  reads
$$
\D=\pmatrix{&T^{\dagger}\cr T&\cr}
=
\pmatrix{&\vec\sigma\cdot\vec\pi-i\textstyle{\lambda\over r}\ccr
\vec\sigma\cdot\vec\pi+i\textstyle{\lambda\over r} &\cr}.
\equation
$$
$\D$ anticommutes with the chirality operator $\gamma^5$.

The square of the Dirac operator,
$$
H =\pmatrix{H_1&\cr &H_0\cr}
=\pmatrix{T^\dagger T&\cr &TT^\dagger\cr}
=
\bigg\{
\vec\pi^2 -q{\vec{\sigma}\cdot\vec{ r}\over r^3} +{\lambda^2\over r^2}
-\lambda\gamma^5{\vec{\sigma}\cdot\vec{ r}\over r^3}
\bigg\},
\equation
$$
is a supersymmetric
Hamiltonian  [8]: the non-zero-energy parts
of the chiral sectors, defined as the eigensectors of $\gamma^5$,
are intertwined by the unitary transformations
$$
U=T\,{1\over\sqrt{H_1}}
\and
U^{-1}=U^{\dagger}=
{1\over\sqrt{H_1}}\, T^{\dagger},
\qquad
UH_1U^\dagger=H_0.
\equation
$$
The partner hamiltonians
$H_1$ and $H_0$, have therefore the same positive spectra, and
the spectrum of $\D$ can be obtained from that of $\D^2$.
For further analysis, it is convenient to set
$$
\sigma_r=\vec{\sigma}\cdot(\vec{ r}/r),
\qquad
z=\vec\sigma\cdot\vec\ell+1 .
\equation
$$
Note that $\sigma_r{}^2=1$, and that $z$
anticommutes with $\sigma_r $ and $\vec\sigma\cdot\vec\pi$,
$\quad
\lbrace z, \sigma_r \, \rbrace = 0
$
and
$\quad
\lbrace z, \vec\sigma\cdot\vec\pi\rbrace=0.
$
Also $z$ is equal to the supercharge $\widetilde{Q}_D$ of (1.2).
For $\lambda\neq0$, $z$ satisfies the relations
$zT=-T^\dagger z$,
$zT^\dagger =-Tz$,
$zH_1=H_0z$
and
$zH_0=H_1z.$
Therefore, the operator
$$
{\cal K}=\pmatrix{&-iz\cr iz&\cr}
\equiv
i\pmatrix{&-\vec\sigma\cdot\vec\ell-1\cr\vec\sigma\cdot\vec\ell+1&\cr}
\equation
$$
commutes with the Dirac operator $\D$,
and hence also with its square.
Using
$(\vec\sigma\cdot\vec L)^2
=
\vec{ L}^2+i\vec{\sigma}\cdot (\vec{ L}\times\vec{ L})
=
\vec{ L}^2 -\vec\sigma\cdot\vec{L}$,
one proves furthermore that
$
{\cal K}^2\,=\,z^2\,=\,\vec{ J}^2 + 1/4-q^2.
$
 Thus, since the eigenvalues of
$\vec{ J}^2$ are
$j(j+1)$, $j=q-1/2,\ q+1/2,\ldots$, the operators
$z$ (and $\cal K$) have {\it irrational} eigenvalues,
$$
\kappa =\pm\sqrt{(j+1/2)^2-q^2}\ .
\equation
$$

The operator ${\cal K}$
is hermitian because $j\geq q-1/2$.
For the lowest allowed value $j=q -1/2$, however,
$\kappa$
vanishes and ${\cal K}$ is not invertible.
The operator ${\cal K}$  has been used by Dirac
in the study of the relativistic hydrogen atom [3]  long time
ago; its form adapted to the monopole,  (2.7), was found by
Berrondo and McIntosh  [5].
It is  more convenient
to use, however,  the hermitian {\it Biedenharn operator} [4]
$$
\Gamma=
-\pmatrix{\vec\sigma\cdot\vec\ell+1+\lambda\sigma_r&
\cr
&\vec\sigma\cdot\vec\ell+1-\lambda\sigma_r
\cr}
\equiv-\pmatrix{y&\ccr&x}.
\equation
$$
Here
$
y=z+\lambda \sigma_r
$
and
$x=z-\lambda \sigma_r,
$
so that $x\sigma_r =- \sigma_r y$.
The eigenvalues of $\Gamma$,
$$
\gamma=\pm\sqrt{(j+1/2)^2+\lambda^2-q^2},
\qquad
(\hbox{sign}\ \gamma=-\hbox{sign}\ \kappa)
\equation
$$
are in general still  irrational.
However, owing to the presence of $\lambda^2$,
the operator $\Gamma$ is invertible
whenever $\lambda^2>0$.
The clue is that, in terms of $\Gamma$, $\D^2$ becomes simply
$$
\D^2\equiv
\pmatrix{H_1\cr &H_0\cr}
=
-(\partial _r + {1\over r})^2+{\Gamma(\Gamma +1)\over r^2}.
\equation
$$
Here $p_r=-i(1/r)\partial_r \, r =-i(\partial_r +1/r)$
is the hermitian operator conjugate to $r$.
This is conveniently checked by writing,
using the radial form
$\vec\sigma\cdot\vec\pi=-i\sigma_r(\partial_r+1/r - z/r)$.
The supercharges $T$ and $T^{\dagger}$ as
$$\matrix{
&T&=&-i\sigma_r \Big(\partial_r+{\displaystyle 1\over\displaystyle r}
- {\displaystyle y\over\displaystyle r}\Big)
&=
&-i\Big(\partial_r+{\displaystyle 1\over\displaystyle r}
+{\displaystyle x\over\displaystyle r}\Big) \sigma_r,
\ccr
&T^{\dagger}&=&-i\sigma_r\Big(\partial_r+{\displaystyle 1\over\displaystyle r}
-{\displaystyle x\over\displaystyle r}\Big)
&= &-i\Big(\partial_r+{\displaystyle 1\over\displaystyle r}
+ {\displaystyle y\over\displaystyle r}\Big)\sigma_r.
\cr}
\equation
$$

The self-adjointness of $\D^2$ requires $\vert\lambda\vert\geq 3/2$ [8].
 It follows from our previous formul{\ae} that
$\Gamma$ anticommutes with $\D$  and commutes therefore
with its square. Hence the shifted operators $x$ and $y$
commute with the partner hamiltonians,
$
[x, H_0]=0=[y, H_1].
$
 In conclusion, our $4$-component operators satisfy the
non-linear algebra
$$\matrix{
\D^2=H,
&\big\{\Gamma,\D\big\}=0,
&\big[\Gamma, H\big]=0,
&\Gamma^2={\vec{J}}\strut^2+1/4+\lambda^2-q^2,
\ccr
\big[\vec{J}, \Gamma\big]=0,
&\big[\vec{J}, \D\big]=0,
&\big[\vec{J}, H\big]=0,
&\big[J_i, J_j\big]=i\epsilon_{ijk}J_k,
\cr}
\equation
$$
which only differs from that found for
the $2$-components objects of
 [1] in the appearance of $\lambda^2$ in $\Gamma^2$.
Note that it is now $\D$ which plays the r\^ole of $Q_D$, and
$\Gamma$ plays that of $\widetilde{Q}_D$.

\goodbreak
\chapter{Explicit solutions}

A look at (2.10) shows that
the Biedenharn operator $\Gamma$ plays clearly a r\^ole
analogous to angular momentum.
Since $\Gamma$ and $J$ commute,
they can be simultaneously diagonalized.
A  convenient basis is found as follows [9].
Let us first assume that
$j\geq q+1/2$,
and let $L_{\pm}\=j\pm1/2$ be the orbital angular momentum quantum
number. Consider first the two-component spinors
$$
\varphi_{\pm}^{\mu}
=\sqrt{{L_{\pm}+1/2\mp\mu\over 2L_{\pm}+1}}\,
Y^{\mu-1/2}_{L_{\pm}}\pmatrix{1\ccr 0\cr}
\mp
\sqrt{{L_{\pm}+1/2\pm\mu\over 2L_{\pm}+1}}\,
Y^{\mu +1/2}_{L_{\pm}}
\pmatrix{0\ccr 1\cr},
\equation
$$
where the $Y$'s are the monopole harmonics, defined in [10].
The $\varphi$'s  satisfy \hfill\break
$
\vec{ J}^2= j(j+1),
\;
J_3 =\mu,\ (\mu = -j,\cdots,j),
\;
\vec{ L}^2= L_{\pm}(L_{\pm}+1),
$
\noindent and the action of $\sigma_r$ upon the $\varphi$'s can be obtained
from [11].
Then, dropping the upper index $\mu$, and using \hfill\break
$\vec\sigma\cdot\vec\ell=
\vec{J}^2-\vec{L}^2-3/4+q\vec\sigma\cdot\vec{r}/r$,
it may be shown that
the $2$-spinors
$$
\chi_{\pm}
=
{q\over j+1/2 +\mid\kappa\mid}\ \varphi_{\pm}
\mp \varphi_{\mp}
\equation
$$
satisfy
$
z\ \chi_{\pm}
=
\pm\vert\kappa\vert\ \chi_{\pm}
$
and
$\sigma_r \ \chi_{\pm}
=
\chi_{\mp},
$
as well as
$\vec{ J}^2=j(j+1)$,
$J_3=\mu,\ (\mu = -j,\cdots, j)$.
Hence, defining $\gamma$ by $\gamma^2=\kappa^2+\lambda^2$, we have
$$
\matrix{
\phi_+=(\vert\kappa\vert+\vert\gamma\vert)\ \chi_+-\lambda\ \chi_-,
&\quad
\phi_-=\,\lambda\ \chi_+ +(\vert\kappa\vert+\vert\gamma\vert)\chi_-
\cr\cr
\Phi_+=(\vert\kappa\vert+\vert\gamma\vert)\ \chi_++\lambda\ \chi_-,
&\quad\Phi_-=-\lambda\ \chi_+ +(\vert\kappa\vert+\vert\gamma\vert)\chi_-
\cr}
\equation
$$
diagonalize $x$ and $y$,
$
x\phi_{\pm}
=
\mp\mid\gamma\mid\phi_{\pm}
$
and
$y\Phi_{\pm}
=
\mp\mid\gamma\mid\Phi_{\pm}.
$
The operator $\sigma_r  =\vec\sigma\cdot{\vec{ r}/r}$ interchanges
the $x$ and $y$ eigenspinors,
$
\sigma_r \,\phi_{\pm}
=
\Phi_{\mp},
$
a result which also follows directly from $x \sigma_r=-\sigma_r y$.

When $j=q-\2$, there are no $L_-=q-1$ states,
though: we only have  $2(q-\2)+1=2q$ states with $L=L_+=q$ , namely
$$
{\varphi}^0_+
=
\sqrt{{q+1/2+\mu\over 2q+1}}\,
Y^{\mu-1/2}_q\pmatrix{1\ccr 0\cr}
+
\sqrt{{q+1/2-\mu\over 2q+1}}\,
Y^{\mu +1/2}_q
\pmatrix{0\ccr 1\cr},
\equation
$$
Thus, for $j=q-\2$, no $\varphi_-$ is available, and
${\chi}^0_+=-{\chi}^0_-={\varphi}^0_+$
is annihilated by $z $.
Therefore, there are no
$\phi_-$-states in the $\gamma^5=-1$ sector, and no $\Phi_+$ states
in the $\gamma^5=1$ sector.
In each $\gamma^5$ sector, (3.4)
yields in turn $(2q)$ states, namely
$$
{\phi}^0_+
=
-{\Phi}^0_-
\;\propto\;
{\varphi}^0_+.
\equation
$$
They are eigenvectors of $x$ and  $y$
with eigenvalues $\pm\lambda$, respectively,
and are still interchanged by  $\sigma_r$.
In conclusion,
the eigenfunctions of $\D^2$ are
$$\matrix{
&\left\{
\eqalign{&
\Psi_{\pm
}
=u_{\pm}
\pmatrix{\Phi_{\pm}\cr 0\cr}
\ccr
&\psi_{\pm
}
=
u_{\pm}
\pmatrix{0\cr\phi_{\pm}}
\cr}
\right.
\qquad\hbox{for}\quad
\left\{\matrix{\gamma^5 = 1\cr\cr\cr
\gamma^5=- 1\cr}\right.
&\hbox{for}\;j\geq q+1/2,
\ccr
&\left\{
\eqalign{&
\Psi_-^0=u_-^0
\pmatrix{\Phi_-\cr 0\cr}
\ccr
&\psi_+^0=u_+^0
\pmatrix{0\cr\phi_+\cr}
\cr}
\right.
\qquad\hbox{for}\quad
\left\{\matrix{\gamma^5 =1\cr\cr\cr
\gamma^5=-1\cr}\right.
&\hbox{for}\;j= q-1/2.
 \cr}
\equation
$$
where $u_\pm$ solves the radial equation
$$
\left[
-\left(\partial_r + {1\over r}\right)^2+{\gamma(\gamma +1)\over r^2}-E
\right]u_\pm(r)=0.
\equation
$$
There are no bound states; the scattering states involve  Bessel functions:
$$
u_\pm(r)={1 \over \sqrt{r}} \, J_{|\gamma+\2|}\big(\sqrt{E}\,r\big).
\equation
$$

\chapter{Symmetries}

A spin $0$ particle in the field of a Dirac monopole has an $\o(2,1)$
symmetry, generated by $H\equiv H_0=\vec\pi^2$
and by dilations and expansions  [12],
$$
D=tH-1/2\big\{\vec\pi\cdot\vec r\big\}
\qquad
K=-\2t^2H+tD+\2r^2.
\equation
$$
This symmetry has been extended to
the Pauli Hamiltonian (1.1) with formally the same
generators (4.1), with $H\equiv H_D$ replacing $H_0$  [2].
The supercharge $Q_D$ is a square-root of $H_D$.
Commuting $Q_D$ with the expansion, $K$, yields a new fermionic
generator, namely
$$
S=i[Q_D, K]={1\over\sqrt{2}}\vec\sigma\cdot\big(\vec r-\vec\pi t),
\equation
$$
and it is then readily proved that the bosonic operators
$H_D$, $D$, $K$ close,
with $Q$ and $S$ into an $\osp(1/1)$ superalgebra.
[2, 8, 13].
Now remarkably
$$
i[Q_D,S]-\2=z,
\equation
$$
and $z^2$ is a Casimir operator of this $\osp(1/1)$ [8].

The same bosonic $\o(2,1)$ symmetry arises
for the generalized monopole system (2.4).
The Dirac operator
$Q\equiv\D$ is a square-root of $H$ by construction. However,
$$
Q^\star=\gamma^5\,Q
\equiv
\pmatrix{&\vec\sigma\cdot\vec\pi-i\textstyle{\lambda\over r}\ccr
-\vec\sigma\cdot\vec\pi-i\textstyle{\lambda\over r} &\cr}.
\equation
$$
is a {\it new} square-root, $\{Q^\star, Q^\star\}=H$.
Commuting  $K$ with $Q$ and with $Q^\star$ yields
$$
S=\gamma^5\vec\gamma\cdot\vec r-tQ
\and
S^\star=-i\gamma^5S.
$$
In this way, we get two, independent, super-extensions
of the bosonic $\o(2,1)$. The two $\osp(1/1)$'s do not close yet:
the ``mixed'' anticommutators between the $Q$-type and $S$-type
charges yield a new bosonic charge, namely
$$
Y=\{Q, S^\star\}=-\{Q^\star, S\}
=\gamma^5\big(
z
+\smallover1/2\big)
-\lambda \sigma_r,
\equation
$$
that commutes with the other bosonic charges.
The four operators $H, D, K, Y$ do close finally with the
four fermionic charges $Q, Q^\star, S, S^\star$,
$$
\matrix{
[Q,D]\hfill&=&iQ,\hfill
&[Q^\star,D]\hfill&=&iQ^\star,\hfill
\ccr
[Q,K]\hfill&=&-iS,\hfill
&[Q^\star,K]\hfill&=&-iS^\star,\hfill
\ccr
[Q,H]\hfill&=&0,\hfill
&[Q^\star,H]\hfill&=&0,\hfill
\ccr
[Q,Y]\hfill&=&-iQ^\star,\hfill
&[Q^\star,Y]\hfill&=&iQ,\hfill
\ccr
[S,D]\hfill&=&-iS,\hfill
&[S^\star,D]\hfill&=&-iS^\star,\hfill
\ccr
[S,K]\hfill&=&0,\hfill
&[S^\star,K]\hfill&=&0,\hfill
\ccr
[S,H]\hfill&=&2iQ,\qquad\hfill
&[S^\star,H]\hfill&=&2iQ^\star,\hfill
\ccr
[S,Y]\hfill&=&-iS^\star,\hfill
&[S^\star,Y]\hfill&=&iS,\hfill
\ccr
\{Q,Q\}\hfill&=&H,\qquad\qquad\hfill
&\{Q^\star,Q^\star\}\hfill&=&H,\hfill
\ccr
\{S,S\}\hfill&=&2K,\hfill
&\{S^\star,S^\star\}\hfill&=&2K,\hfill
\ccr
\{Q,Q^\star\}\hfill&=&0,\hfill
&\{S,S^\star\}\hfill&=&0,\hfill
\ccr
\{Q,S\}\hfill&=&-D,\hfill
&\{Q^\star,S^\star\}\hfill&=&-D,\hfill
\ccr
\{Q,S^\star\}\hfill&=&Y,\hfill
&\{Q^\star,S\}\hfill&=&-Y.\hfill
\cr}
\equation
$$
which are the commutation relations of the $\osp(1/2)$
superalgebra,
to which spin adds an extra $\o(3)$ [8].
Now the Casimir of $\osp(1/2)$ is the square of
$$
i[Q, S]-\2=i[Q^\star, S^\star]-\2=\Gamma,
\equation
$$
which provides a nice  interpretation
for  the Biedenharn operator $\Gamma$.
Similar algebras were studied in [14].

\chapter{Particular cases}

(i)  For $\lambda\=0$ , we have $Q=Q^\dagger=Q_D$,
$H_1=H_0=H_D$,  the Pauli Hamiltonian in a pure monopole field  [2].
The $4$-component Hamiltonian is simply {\rm diag}$(H_D, H_D)$;
the Biedenharn and the Dirac operators are related as
$\Gamma=-i\gamma^4{\cal K}$.
In this case,
we recover the formul{\ae} in [1, 2] .

(ii) Another particular value is $\lambda\=\pm q$, when
the situation is similar to that
in Taub-NUT space [15]:
the spin drops out in one of the chiral sectors,
while the Pauli term gets doubled in the other.
For $\lambda=q$, for example, the Hamiltonian
(2.4) reduces to
$$
H=\pmatrix{H_1&\ccr &H_0\cr}=
\pmatrix{
\vec\pi^2
 +\displaystyle{q^2\over r^2}
  -
2q\displaystyle{\vec\sigma\cdot\vec{ r}\over z\displaystyle r^3} &
 \ccr
&\vec\pi^2
+\displaystyle{q^2 \over r^2}\cr
}.
\equation
$$
Here $H_0$ describes a spin $0$ particle in the combined  field of
a Dirac monopole and of an inverse-square potential, while
$H_1$ corresponds to a particle with anomalous gyromagnetic ratio $4$.
The Biedenharn operator has half-integer eigenvalues, $\gamma=\pm (j+\2)$.
Note that the $\gamma^5=-1$ eigenspinors now reduce to those in Eq. (3.1),
$
\phi_\pm
\;\propto\;
\varphi_\pm.
$

Assume first that $j\geq q+1/2$.
Since $\gamma(\gamma+1)$ is now the same for
$-\vert\gamma\vert$ as for $\vert\gamma\vert -1$,
these values lead to identical solutions.
Therefore, in each $\gamma^5$ sector,
the  energy levels are two-fold degenerate.
The numerator of the $r^{-2}$ term  reads
$$
\matrix{(j+\smallover3/2)(j+\smallover1/2)&
\ccr
&(j+\smallover1/2)(j-\smallover1/2)
\cr}
\qquad\hbox{for}\;
\left\{\matrix{\gamma>0\ccr\cr\gamma<0\cr}\right.
\equation
$$
so that we indeed get the same equation with $\gamma>0$  for $j$,  as with
$\gamma<0$ for $(j-1)$, provided that $(j-1)$ states do exist.
In both cases, (3.8) yields
$
r^{-1/2}\, J_{j+1}\big(\sqrt{E}\,r\big).
$

The two-fold degeneracy is hence also explained
by  an extra $\o(3)$ symmetry in addition to the rotational symmetry:
 spin is trivially conserved for $H_0$, and this is exported to
$H_1$  by supersymmetry. The extra $\o(3)$ symmetry is
generated hence by the spin vectors
$$
\vec{S}_0= \2\vec{\sigma} \qquad \hbox{for} \;H_0 ,
\quad
\vec{S}_1=U^{\dagger}\vec{ S}_0U \qquad\hbox{for} \;H_1 ,
\equation
$$
where $U$ and  $U^{-1} = U^{\dagger}$
are the intertwiners of (2.4).
The two-fold degeneracy corresponds precisely to this $\o(3)$ symmetry.
For $j\=q-1/2$ , half of the states are missing.

The system admits further symmetries. Firstly,
$H_0$ admits the
non-relativistic conformal $\o(2,1)$ symmetry [10];
supersymmetry
exports this to the partner Hamiltonian $H_1$. The symmetries
combine with $\D$ and $-i\gamma^5\D$
into an $\osp(2,1)$ superalgebra [8] .

(iii) Replacing the scalar potential
$\lambda/r$ by $q(1-1/r)$ ---
which corresponds to the long-range
limit of the scalar field  of a self-dual ``BPS'' monopole [14] ---
we get
$$
\pmatrix{H_1&\cr&H_0\cr}
=
\pmatrix{
\vec{\pi}^2
-\displaystyle{2q^2 \over r}
+\displaystyle{q^2\over r^2}
+q^2
-2q\displaystyle{\vec{\sigma}\cdot\vec{r}\over r^3}
&
\ccr
&
\vec{\pi}^2
-\displaystyle{2q^2 \over r}
+\displaystyle{q^2\over r^2}+ q^2
\cr}.
\equation
$$

The ``lower'' Hamiltonian,
 $H_0$, has again gyromagnetic ratio $0$ yielding an extra
 $\o(3)$ symmetry. Its properties have been thouroughly
 studied by McIntosh and Cisneros, and by Zwanziger [17], who
 found that it admits
 a Kepler-type dynamical symmetry.
 Its superpartner $H_1$ describes a spinning particle
of anomalous gyromagnetic ratio $4$: this is  the
 \lq dyon' of D'Hoker and Vinet [18].
Then supersymmetry can be used to transfer the symmetries of
$H_0$ to $H_1$ [16]; the system can  be solved using the
Biedenharn method [9].
\goodbreak

\chapter{Minkowski space extension}

\vskip 1pt
In this section we turn to the Dirac-equation in Minkowski space-time,
with a combined Coulomb-monopole and massless scalar background.
This relativistic system of equations has properties very similar to
the Euclidean Dirac problem considered above, and can be solved by
analogous methods.

In Minkowski space-time, with metric $\eta_{\mu \nu} =$
diag$(-1,+1,+1,+1)$, we use  Dirac matrices with the
properties $ \{ \gamma_\mu \, , \, \gamma_\nu \} = 2\eta_{\mu \nu}$ and
$\gamma_\mu{}^\dagger =-\gamma_0 \gamma_\mu \gamma_0 $.

Our starting point is the Dirac equation
$$ (i\D+m+g\varphi)\psi=0 \quad . \eqno(6.1) $$
\noindent
Here $\varphi=-\tilde g /r$ is a dynamical scalar background and
$m$ the mass (which
can be taken as the vacuum expectation value of the scalar field). Eq. (6.1)
describes the motion in the long-range field of a Julia-Zee dyon [19].
The covariant derivatives contain electromagnetic background
potentials corresponding to a Coulomb field
$$ D_0=\partial_0+iq\phi \quad , \quad \phi=\tilde q /r \quad ,  $$
and the field of a magnetic monopole taken, as in Sec. 2, to be of
unit strength.

Multiplication by $(-i \D + m + g \varphi)$ gives a Klein-Gordon type
equation, which on stationary states $\psi({\bf r},t) = \exp(- iEt)
\psi_E({\bf r})$ takes the form
$$
\left[ - (E - q \phi)^2 - (\vec{\nabla} - iq \vec{A})^2 +
(m + g \varphi)^2 - iq \sigma^{\mu\nu} F_{\mu\nu} -
ig \gamma_\mu \partial^\mu  \, \varphi \right]\,
\psi_E\, =\, 0.
\eqno(6.2) $$

The generalized Klein-Gordon operator can be block-diagonalized in
$(2 \times 2)$ blocks, with 2-component eigenspinors $\psi_{\pm}$
satisfying
$$ \left[ - (E - q \phi)^2 - (\vec{\nabla} - iq \vec{A})^2 +
(m + g \varphi)^2 -
\vec{\sigma} \cdot ( q \vec{B} \pm
\vec{\nabla} \Lambda ) \right] \, \psi_{E,\pm}\, =\, 0,
\eqno(6.3) $$
with $ \Lambda \, =\, \sqrt{g^2 \varphi^2 - q^2 \phi^2}$.
Note that the square root is real for $g^2 \varphi^2 \geq q^2 \phi^2$,
and imaginary for $g^2 \varphi^2 < q^2 \phi^2$. For the case of the
Coulomb and scalar potentials  this becomes
$$
\Lambda \, =\, {\lambda \over r} \quad , \quad
 \lambda \, =\, \sqrt{g^2 \tilde{g}^2 - q^2 \tilde{q}^2}.
\eqno(6.4) $$

Defining the operators $\vec{\ell}$ and $\vec{J}$ as in Sec. 2, we may
cast (6.3) into the form
$$
 \lbrack - ( \partial_r + {\smallover 1/r} )^2  +
{1 \over r^2} (  \vec{J}^2 -
{\smallover 3/4} -\vec{\ell}\cdot\vec{\sigma}-
q^2 \mp \lambda  \sigma_r )
-( E -
 {{q \tilde q} \over r} )^2 + ( m - {{g \tilde g} \over r}
 )^2 \rbrack \psi_{E,\pm} = 0.
\eqno(6.5) $$
To make contact with the Biedenharn operator and the work of Sec. 2, we note
that the operators $y$ and $x$ of (2.8) occur as blocks in (6.5). Writing here
$\Gamma_+=-y$ and $\Gamma_-=-x$, we have
$$
\Gamma_{\pm}\, \left( \Gamma_{\pm} + 1
\right) \, = \vec{\ell}^2 \,+\, \lambda^2 \, -\, \left( q
 \pm \lambda \right) \, \sigma_r \quad ,
\eqno(6.6) $$
and can hence rewrite (6.5) in the standard form
$$
\left[ - \left( \partial_r + {1 \over r} \right)^2\, +\, {1 \over r^2}\,
 \Gamma_{\pm} \left( \Gamma_{\pm} + 1 \right) \, +\, {2k \over r}\,
 +\, \varepsilon \right] \, \psi_{E,\pm}\, =\, 0 \quad .
\eqno(6.7) $$
Here the constants $k$ and $\varepsilon$ are given by
$
k\, =\, q\tilde{q} E\, -\, g\tilde{g} m$ and
$\varepsilon \, =\, m^2\, -\, E^2$.
Note the symmetry under the simultaneous exchange of
$(E,q\tilde{q}) \leftrightarrow i(m,g\tilde{g})$. The
eigenvalues of $\Gamma_{\pm}$ are the ones given in (2.10): here
$$
\gamma\, =\, \pm\, \sqrt{ \left( j + {\smallover1/2} \right){}^2 +
 g^2 \tilde{g}^2 - q^2 (\tilde{q}^2 + 1)}\quad .
\eqno(6.8) $$
Introducing the notation $l_\gamma$, where $l_\gamma=\gamma$ for
$\gamma \geq 0$ and $l_\gamma=-(1+\gamma)$ for $\gamma<0$,
we see that the eigenstates of $\Gamma_{\pm}$ satisfy the equation
$$
\left[ - \left( \partial_r + {1 \over r} \right)^2\, +\, {1 \over r^2} \,
 l_{\gamma} \left( l_{\gamma} + 1 \right) \, +\, {{2k} \over r}\,
 +\, \varepsilon  \right]\, \psi_{E,l(\gamma)}\, =\, 0.
\eqno(6.9) $$
The spectrum of eigenvalues for bound states is well-known
from atomic physics:
$$
\varepsilon = k^2 \, /{n^2_\gamma} , \qquad
n_{\gamma} = 1 + l_{\gamma} + N, \qquad  N = 0,1,2,...\, .
\eqno(6.10) $$
However, in this case the bound-state energy eigenvalues
themselves then are given by
$$
E(j,N)\, =\, m\, \left( {{ g\tilde{g}q\tilde{q}} \pm
 n_{\gamma} \sqrt{n_{\gamma}^2  + q^2 \tilde{q}^2 - g^2 \tilde{g}^2}}
\right) \quad .
\eqno(6.11) $$

For the ground state $j = 0$, $N = 0$, the wave equation
factorizes as expected on the basis of supersymmetric
quantum mechanics:
$$
\left( - \partial_r + {{\gamma_g - 1} \over r} + {k_g \over \gamma_g}\right)
\,  \left( \partial_r + {{\gamma_g +1} \over r} + {k_g \over \gamma_g}
\right)\,
 \psi_0\, =\, 0 \quad , \eqno(6.12)
$$
with $\gamma_g$ and $k_g$ the ground state values of $\gamma$ and $k$.

\goodbreak
\kikezd{Acknowledgements.}
 P. A. H. and A. J. M would like to thank NIKHEF for
 the hospitality extended,
 while part of this work was completed.
 Correspondence with Dr. M. Plyushchay
 is also acknowledged.

\vskip3mm\goodbreak

\centerline{References}

\reference 
F. De Jonghe, A. J. Macfarlane, K. Peeters, J.-W. van Holten,
{\sl Phys. Lett.} {\bf B 359}, 114 (1995).

\reference 
E. D'Hoker and L. Vinet,
{\sl Phys. Lett}. {\bf B 137}, 72 (1984).

\reference 
P. A. M. Dirac,
{\it The Principles of Quantum Mechanics}, Oxford : Clarendon (1958).

\reference 
L. C. Biedenharn, {\sl Phys. Rev}. {\bf 126}, 845 (1962);
L. C. Biedenharn and N. V. V. Swamy, 
{\sl ibid}. {\bf 5B}, 1353 (1964).

\reference 
M. Berrondo and H. V. McIntosh,
{\sl Journ. Math. Phys}. {\bf 11}, 125 (1970).

\reference 
M. Plyushchay,
hep-th/0005122, to be published in Phys. Lett. {\bf B}.

\reference 
D. Spector, {\sl Phys. Lett.} {\bf 474B}331-335 (2000).

\reference
E. D'Hoker and L. Vinet,
{\sl Comm. Math. Phys}. {\bf 97}, 391-427 (1985).

\reference 
F. Bloore and P. A. Horv\'athy,
{\sl Journ. Math. Phys}. {\bf 33}, 1869 (1992).

\reference 
T. T. Wu and C. N. Yang, {\sl Nucl. Phys.} {\bf B107} 365-380 ( 1976).

\reference 
Y. Kazama, C. N. Yang and A. S. Goldhaber, {\sl Phys. Rev.} {\bf 15D}
2287-2299 (1977).

\reference 
R. Jackiw,
{\sl Ann. Phys} (N.Y.) {\bf 129}, 183 (1980).

\reference 
C. Duval and P. A. Horv\'athy,
{\sl Journ. Math. Phys.} {\bf 35}, 2516 (1994).

\reference 
K. S. Nirov and M. Plyushchay,
{\sl Phys. Lett}.  {\bf B405}, 114 (1997).

\reference 
M. Visinescu,  {\sl Phys. Lett.} {\bf B 339}, 28 (1994);
J.-W. van Holten,
{\it ibid}.  
{\bf B 342}, 47 (1995);
A. Comtet and P. A. Horv\'athy,
{\it ibid}.  
{\bf B 349}, 49 (1995).

\reference 
L. Gy. Feh\'er, P. A. Horv\'athy and L. O'Raifeartaigh,
{\sl Int. Journ. Mod. Phys}. {\bf A4}, 5277 (1989).

\reference 
H. V. McIntosh and A. Cisneros,
{\sl Journ. Math. Phys}. {\bf 11}, 896 (1970);
D. Zwanziger, {\sl Phys. Rev.} {\bf 176}, 1480 (1968).

\reference 
E. D'Hoker and L. Vinet,
{\sl Phys. Rev. Lett}. {\bf 55}, 1043 (1986).

\reference 
B. Julia and A. Zee, {\sl Phys. Rev.} {\bf D11} 2227 (1975).

\bye